\documentclass[11pt, letterpaper]{article}
\usepackage[letterpaper,margin=1in]{geometry}
\usepackage{amsmath}
\usepackage{amsthm}
\usepackage{amssymb}
\usepackage{thmtools}
\usepackage{thm-restate}
\usepackage{mathtools}
\usepackage{setspace}
\usepackage{algorithm}
\usepackage{times}
\usepackage{complexity}

\usepackage[table]{xcolor}

\definecolor{linkcol}{rgb}{0,0,0.38}
\definecolor{citecol}{rgb}{0.8,0,0}
\definecolor{urlcol}{rgb}{0.1,0.35,0}

\usepackage[pdfpagelabels,bookmarks=false,hyperfootnotes=false]{hyperref}
\hypersetup{colorlinks, linkcolor=linkcol, citecolor=citecol, urlcolor=urlcol}

\usepackage[utf8]{inputenc}

\usepackage[
backend=biber,
style=alphabetic,
citestyle=alphabetic,
maxalphanames=4,
maxcitenames=99,
mincitenames=98,
maxbibnames=99,
giveninits=true,
backref=true,
]{biblatex}

\usepackage[nameinlink]{cleveref}

\usepackage{bbm}
\usepackage{thm-restate}

\newtheorem{theorem}{Theorem}[section]
\newtheorem{lemma}[theorem]{Lemma}

\newtheorem{proposition}[theorem]{Proposition}
\newtheorem{definition}[theorem]{Definition}
\newtheorem{remark}[theorem]{Remark}
\newtheorem{corollary}[theorem]{Corollary}

\usepackage{url}
\usepackage{array}
\usepackage{setspace}
\usepackage{wrapfig}

\usepackage[inline]{enumitem}
\usepackage{bm}
\usepackage{ifthen}

\setlist[enumerate]{sep=0pt,topsep=0.5em}
\setlist[enumerate,1]{label=(\roman*), leftmargin=2.2em}
\setlist[itemize]{nosep,topsep=0.3em}

\usepackage[textsize=footnotesize, color=blue!30!white]{todonotes}
\usepackage{xfrac}
\usepackage{tikz}
\usetikzlibrary{calc}
\usetikzlibrary{math}
\usetikzlibrary{shapes.geometric}
\usetikzlibrary{arrows}
\usetikzlibrary{decorations.pathreplacing, decorations.pathmorphing}

\usepackage{tcolorbox}
\tcbuselibrary{skins}

\usepackage{float}
\usepackage{graphicx}
\usepackage[margin=10pt,font=small,labelfont=bf,skip=-5pt]{caption}
\usepackage{subcaption}

\usepackage[linesnumbered,vlined,ruled,algo2e]{algorithm2e}
\SetAlgoSkip{bigskip}
\SetAlgoInsideSkip{smallskip}

\renewcommand{\epsilon}{\varepsilon}

\newcommand{\DBF}{\textsf{Degree-Bounded Forests}}
\newcommand{\DBFs}{\textsf{DBF}}

\newcommand{\cI}{\mathcal{I}}
\newcommand{\bmin}{b_{\min}}
\newcommand{\cM}{\mathcal{M}}

\title{Approximation Algorithms for Maximum-Weight Degree-Bounded Forests}
\author{Theophile Thiery\thanks{
Department of Computer Science, ETH Zurich, Zurich, Switzerland.
Email: \href{mailto:theophile.thiery@inf.ethz.ch}%
{theophile.thiery@inf.ethz.ch}.}
\and
Vera Traub\thanks{
Department of Computer Science, ETH Zurich, Zurich, Switzerland.
Email: \href{mailto:vtraub@ethz.ch}%
{vtraub@ethz.ch}.}
\and
Nicola Widmer
\thanks{
Department of Computer Science, ETH Zurich, Zurich, Switzerland.
Email: \href{mailto:widmenic@student.ethz.ch}%
{widmenic@student.ethz.ch}.}
}

\date{}

\begin{document}

\maketitle

\begin{abstract}
    In the maximum-weight degree-bounded forest problem, we are given a graph with nonnegative edge weights and a degree upper bound on each vertex. The objective is to find a maximum-weight forest satisfying the degree bounds.
    We show that this problem admits a simple $2$-approximation via the greedy algorithm. This already improves over the best-known guarantees obtained via a reduction to weighted $3$-matroid parity, and sets a baseline to improve upon.
    
    Our main result is an $\sfrac{8}{5}$-approximation algorithm for the maximum-weight degree-bounded forest problem. For the special case of uniform degree bounds, our algorithm achieves an approximation ratio of $\sfrac{3}{2}$.
    It computes two sets of edges, each of which we then restrict to a feasible solution in an optimal way, and finally returns the better of the two resulting solutions.
    To obtain the initial two edge sets, we first relax the degree bounds and leverage an algorithm by Lau and Singh [JACM'15] for degree-bounded spanning trees, and second we relax the forest constraints and employ an algorithm for maximum-weight $b$-matchings. 
\end{abstract}


\thispagestyle{empty}
\addtocounter{page}{-1}
\newpage

\section{Introduction}
    The minimum-cost degree-bounded spanning tree problem is a seminal problem in combinatorial optimization and network design. Given a weighted graph $G = (V, E, w)$ and degree upper bounds $b_v \in \mathbb{N}$ for each vertex $v$, it asks to find a minimum-cost spanning tree satisfying the given degree bounds.
    This line of work culminated in Lau and Singh's iterative relaxation method \cite{Singh-Lau:2015:Approximating} giving a polynomial time algorithm to obtain a spanning tree violating the degree bounds by an additive $1$ while having a cost no more than that of an optimal solution (without violation). Their result is essentially the best possible. 
    The iterative relaxation method was later applied and generalized to variants of this problem, for instance, higher connectivity \cite{Lau-Naor-Salavatipour:2009:Survivable}, directed graphs \cite{Bansal-Khandekar-Nagarajan:2008:Additive}, and matroidal constraints \cite{Kiraly-Lau-Singh:2012:Degree, Zenklusen:2012:Matroidal}.
    
    One downside of these results is that they only obtain a bicriteria approximation, allowing solutions to violate the degree bounds. In this paper, we take an alternative viewpoint. 
    We consider the problem of finding a maximum-weight forest $F \subseteq E$ satisfying all vertex degree bounds. 
    \begin{definition}
    An instance of $\DBF$ ($\DBFs$) consists of a graph $G=(V,E)$ with nonnegative edge weights $w\colon E \to \mathbb{R}_{\geq 0}$ and degree bounds $b\colon V \to \mathbb{N}$.
    The task is to find a maximum-weight forest $F\subseteq E$ satisfying $\deg_F(v) \leq b_v$ for all $v\in V$.
    \end{definition}
    Edge weights are nonnegative without loss of generality, as negative-weight edges cannot be part of an optimal solution.
    This problem is a natural variation of the minimum-cost degree-bounded spanning tree problem.
    Moreover, it is interesting for its combinatorial structure, as it is a special case of weighted $3$-matroid intersection when $G$ is bipartite and of matroid $3$-parity for arbitrary graphs \cite{Bai-Berczi-Siemelink:2025:Approximating}. 

    Due to its connection to multi-matroid optimization problems, it admits a $2$-approximation when $G$ is bipartite via a reduction to $3$-matroid intersection \cite{Lee-Sviridenko-Vondrak:2010:Submodular,Linhares-Olver-Swamy-Zenklusen:2020:Approximate}, and a $4\ln(2) + \varepsilon > 2.772$-approximation for arbitrary $G$ via a reduction to matroid $3$-parity \cite{Feldman-Ward:2026:Submodular}. 
    In the unweighted special case ($w_e=1$ for all $e\in E$), the natural local search algorithm has an approximation ratio of $\sfrac{3}{2} + \varepsilon$ \cite{Lee-Sviridenko-Vondrak:2013:Matroid}. 

    The $\DBF$ problem is also related to $(1, 2)$-TSP which is a notable special case of the traveling salesman problem.
    The $(1, 2)$-TSP problem asks to find a minimum cost tour in a complete graph $G$ where the edge costs are either $1$ or $2$. 
    Any $\alpha$-approximation for $\DBFs$ yields a $(2 - \frac{1}{\alpha})$-approximation for $(1, 2)$-TSP. \footnote{To see this, let $E_1$ be the edges of cost $1$. 
    Any tour restricted to $E_1$ is a \emph{linear forest}, i.e., where components are paths, which corresponds to a forest satisfying degree bounds of $2$ at every vertex. Additionally, any linear forest on $E_1$ of $x$ edges can be extended into a tour of cost $2|V| - x$.  Finding a maximum-size linear forest in $E_1$ is a $\DBFs$ problem, where $w_e = 1$ for all $e \in E_1$ and $b_v = 2$ for all $v\in V$.}
    Since $(1, 2)$-TSP is $\NP$-hard to approximate within a factor $\sfrac{535}{534}$ \cite{Karpinski-Schmied:2012:Approximation}, this implies that $\DBFs$ is $\NP$-hard to approximate with a factor $\sfrac{534}{533} < 1.0019$.

    \subsection{Our Contributions}
    We start by analyzing the greedy algorithm -- a standard heuristic for matroidal problems. We prove that greedy, which in each iteration adds an edge of maximum weight maintaining feasibility, is a $2$-approximation algorithm.
    \begin{restatable}{theorem}{greedy}
        \label{thm:greedy-apx}
        The greedy algorithm is a $2$-approximation algorithm for $\DBF$.
    \end{restatable}
    This already improves over known approximation guarantees in the general case.
    Moreover, even in the bipartite special case it provides a much simpler algorithm than existing ones, while matching the approximation guarantee.
    Not only the algorithm, but also our analysis is relatively simple. We show that the set of feasible degree-bounded forests is a special case of a so-called \emph{$2$-system}. By Korte and Hausmann \cite{Korte-Hausmann:1978:Analysis}, this implies that greedy has an approximation ratio equal to $2$.
    Informally, a $p$-system is a combinatorial structure where a \emph{maximum-sized} solution is at most $p$ times the size of an (inclusion-wise) \emph{maximal} solution. 
    For full details of the proof of \Cref{thm:greedy-apx}, we refer to \Cref{sec:greedy}.

    Our second result is an improvement over the greedy algorithm. We design an $\sfrac{8}{5}$-approximation algorithm for the maximum-weight degree-bounded forest problem $(\DBFs)$. 
    \begin{theorem}
        There is a polynomial-time $8/5$-approximation algorithm for $\DBF$. 
    \end{theorem}
    We complement this result by providing a family of examples where our algorithm attains a ratio arbitrarily close to $\sfrac{8}{5}$.
    In the uniform case, where $b_v = \Delta$ for all $v \in V$ and some $\Delta \geq 1$, we obtain an even better approximation factor of $\frac{3}{2}$.
    More precisely, the approximation factor is $\min\{ \frac{\Delta +1}{2}, \frac{\Delta + 1}{\Delta}\} \leq \frac{3}{2}$ which is tight when $\Delta = 2$; see \Cref{thm:unif-2-3}.

    Moreover, our techniques also imply an upper bound on the integrality gap of a natural linear programming relaxation; see \Cref{sec:integrality_gap}.
    
    To obtain these results, we provide two algorithms performing better than greedy in complementary regimes. 
    First, we observe that the problem is easy when the input graph $G$ is a tree. Indeed, because the graph is acyclic, one can simply compute an optimal $b$-matching (see \Cref{def:b-matching} and \Cref{thm:b-matching}). 
    This invites us to apply Lau and Singh's algorithm \cite{Singh-Lau:2015:Approximating} to obtain a maximum-weight forest $T$ violating the vertex degree bounds by $+1$ and then compute an optimal solution $F_1 \subseteq T$. Importantly, we apply Lau--Singh's algorithm with modified edge weights $w_e' \coloneqq w_e \cdot \frac{\bmin(e)}{\bmin(e) + 1}$ where $\bmin(e) \coloneqq \min \{b_u; b_v\}$ for all $e = (u, v) \in E$. This allows us to show that $w(F_1) \geq w'(F^\ast)$, where $F^\ast$ is an optimal degree-bounded forest. 
    In particular, this procedure has then the same performance as greedy and yields improved guarantees when the minimum degree bound is at least $2$.     
    
    Secondly, we observe that if $G$ already respects the vertex degree bounds, then the problem can be solved optimally by computing a maximum-weight spanning tree.
    A natural algorithm is then to compute a maximum-weight $b$-matching $M$ and then find an optimal solution $F_2 \subseteq M$.
    In fact, we compute a maximum-weight $b$-matching with respect to modified edge weights $w''_e \coloneqq \frac{1}{\bmin(e)} \cdot w_e$.
    Again, we show that $w(F_2) \geq w''(F^\ast)$. 
    This procedure then performs well when the degree bounds are small.
    Returning the maximum weight forest out of these two algorithms yields the desired result.

    For a discussion of why it is crucial to work with the modified weight functions, we refer to \Cref{sec:improving_on_greedy}.
    
    \subsection{Related Work}
        The minimum-cost degree-bounded spanning tree problem is a seminal problem in combinatorial optimization and network design.
        In the special case where the graph is unweighted, Fürer and Raghavachari \cite{Furer-Raghavachari:1994:Approximating} gave an algorithm that returns a spanning tree in which the degree of each vertex $v$ is at most $b_v+1$ (whenever the instance is feasible).
        The first algorithm to output a minimum-cost spanning tree with constant additive degree violation was designed by Goemans \cite{Goemans:2006:Minimum} with $+2$ degree violation. Building on the iterative rounding technique due to Jain \cite{Jain:2001:Factor}, Lau and Singh \cite{Singh-Lau:2015:Approximating} developed an iterative relaxation algorithm that returns a spanning tree with $+1$ violation of the degree bounds.
        We refer the reader to \cite{Gupta-Konemann:2011:Survey, Lau-Ravi-Singh:2011:Iterative, Singh-Lau:2015:Approximating} and references therein for further applications of the iterative rounding method and the iterative relaxation technique.

        The maximum-weight degree-bounded forest problem considered in this paper is related to matroidal problems in the following way. It reduces to the \emph{maximum-weight degree-bounded matroid independent set problem} (\textsf{DBMIS}) introduced by Bai, B\'erczi and Siemelink \cite{Bai-Berczi-Siemelink:2025:Approximating} when $\Delta = 2$. Given a matroid $\cM = (V, \cI)$ where each element has a non-negative weight, a hypergraph $\mathcal{H}$ on the same ground set with maximum degree $\Delta$ and bounds $\{g_e\}_{e \in E(\mathcal{H})}$ on the hyperedges, it asks for a maximum weight independent set $I \in \mathcal{I}$ such that $|I \cap e| \leq g_e$ for all $e \in E(\mathcal{H})$. They show that the feasible sets of \textsf{DBMIS} form a $(\Delta+1)$-extendible system and that the problem reduces to matroid $(\Delta+1)$-parity.
        The matroid $k$-parity problem admits a $\min\{(k+1)\ln(2) + \varepsilon; \frac{k}{2}+o(k)\}$-approximation \cite{Feldman-Ward:2026:Submodular, Singer-Thiery:2026:Submodular}, and a $(\frac{k}{2}+ \varepsilon)$-approximation in the unweighted setting \cite{Lee-Sviridenko-Vondrak:2013:Matroid}. 
        Moreover, for any sufficiently large constant $k \geq f(\varepsilon)$, no polynomial-time algorithm for matroid $k$-parity has an approximation ratio better than $\frac{k}{8 + \varepsilon}$ \cite{Minzer-Zheng:2026:Near} unless $\NP \subseteq \BPP$.
        Analogously to the minimum-cost degree-bounded spanning tree problem, Kir{\'{a}}ly, Lau and Singh \cite{Kiraly-Lau-Singh:2012:Degree} designed an algorithm returning a min-cost \emph{basis} of the degree-bounded matroid independent set problem violating the hyperedges bound by an additive factor $\Delta-1$.

        The degree-bounded forest problem is a constrained forest problem. Another closely related example is the maximum-weight properly colored forest problem (\textsf{PCF}) considered in \cite{Bai-Berczi-Csaji-Schwartz:2026:Approximating}. 
        Given an edge-colored graph, the goal is to find a forest such that any two adjacent edges have distinct colors. For the unweighted problem, they obtained a $9/5$-approximation which was improved to $3/2+ \varepsilon$ via a reduction to $\textsf{DBMIS}$ \cite{Bai-Berczi-Siemelink:2025:Approximating}. 
        \textsf{DBF} with uniform degree bounds $\Delta$ is a special case of \textsf{PCF}. Given an instance $(G,w,\Delta)$ of \textsf{DBF}, we obtain an instance $(G', w)$ of \textsf{PCF} by taking $\Delta$ copies of each edge colored by different colors from $[\Delta]$. Any properly colored forest is degree-bounded with degree at most $\Delta$.
        Finding a properly colored \emph{tree} was considered in \cite{Borozan-Vega-Manoussakis-Martinhon-Muthu-Pham-Saad:2019:Maximum}, investigating its existence from a graph-theoretic and algorithmic perspective. We refer the reader to \cite{Bai-Berczi-Csaji-Schwartz:2026:Approximating, Borozan-Vega-Manoussakis-Martinhon-Muthu-Pham-Saad:2019:Maximum} and references therein for further specific related works and connections to the degree-bounded spanning tree problem.
\section{Analysis of the Greedy Algorithm}\label{sec:greedy}

In this section, we give a simple proof that the greedy algorithm has an approximation ratio of $2$ and that this factor is tight.
The greedy algorithm starts with an empty solution $F=\emptyset$ and iteratively chooses a maximum-weight edge $e$ such that $F\cup \{e\}$ is still a feasible solution.
It then adds $e$ to $F$ and iterates until no such edge $e$ exists anymore.

\greedy*

Our proof shows that the ratio between the \emph{size} of an (inclusion-wise) maximal degree-bounded forest solution to that of a maximum degree-bounded forest is at most $2$. We will then conclude the proof using a result of Korte and Hausmann \cite{Korte-Hausmann:1978:Analysis}: They
proved that the approximation ratio of the greedy algorithm for an arbitrary weighted independence system is equal to the ratio of a maximal base to that of a maximum base; see \Cref{thm:greedyindependence}. We bound this ratio for our problem first.

\begin{lemma}\label{lem:greedyratio}
    Given a graph $G=(V,E)$ and a degree bound $b\colon V\to \mathbb{N}$, let $A,B\subseteq E$ be two inclusion-wise maximal forests satisfying the degree bounds $b$. Then
    \begin{equation*}
        |B|\leq 2|A|.
    \end{equation*}
\end{lemma}

\begin{proof}[Proof of \Cref{lem:greedyratio}]
    Let $S \coloneqq \{v\in V \mid \deg_A(v) = b_v\}$ be the set of vertices for which the degree bounds in forest $A$ are tight, and refer to these as tight vertices. Let $\overline{S} \coloneqq V\setminus S$ be the set of non-tight vertices.

    We partition the forest $B$ into two disjoint sets $B_{S}$ and $B_{\overline{S}}$. The set $B_{S} \subseteq B$ contains the edges incident to at least one tight vertex. The set $B_{\overline{S}} \subseteq B$ contains the edges where both endpoints are non-tight.
    We will show that $|B_{S}|\leq \sum_{v\in S}\deg_A(v)$ and $|B_{\overline{S}}|\leq \sum_{v\in \overline{S}}\deg_A(v)$. From these inequalities, we can then conclude
    \begin{equation*}
        |B| \ =\ |B_{S}| + |B_{\overline{S}}| \ \leq\ \sum_{v\in S}\deg_A(v) + \sum_{v\in \overline{S}}\deg_A(v) \ =\ 2|A|.
    \end{equation*}

    First, we show that
    \begin{equation*}
        |B_{S}| \ \leq\ \sum_{v\in S}b_v \ =\ \sum_{v\in S}\deg_A(v).
    \end{equation*}
    The inequality holds because we can charge each edge in $B_{S}$ to an incident tight endpoint. Since $B$ respects the degree bounds $b$, each vertex $v$ is charged at most $b_v$ times. The equality holds by the definition of the set $S$ of tight vertices.
    
    Next, let $\tilde{S} \subseteq \overline{S}$ be the vertices incident to at least one edge from $B_{\overline{S}}$.
    Consider an $e = \{u, v\} \in B_{\overline{S}}$. We claim that $u$ and $v$ lie in the same connected component of the graph $(V, A)$. Otherwise, the set $A \cup \{e\}$ would be a forest that doesn't violate the degree bounds, contradicting the maximality of $A$.
    Therefore, we have $\deg_A(u) \geq 1$ and $\deg_A(v) \geq 1$.
    We conclude that every vertex from $\tilde{S}$ has degree at least one in the forest $A$.
    Because $B_{\overline{S}}$ is a forest with vertex set $\tilde{S}$, this implies
    \begin{align*}
        |B_{\overline{S}}| &
        \ <\  |\tilde{S}|
        \ \leq\  \sum_{v\in \tilde{S}}\deg_A(v)
        \ \leq\  \sum_{v\in \overline{S}}\deg_A(v). \qedhere
    \end{align*}
\end{proof}

We now state a result by Korte and Hausmann \cite{Korte-Hausmann:1978:Analysis} relating the rank quotient of an independence system to the approximation guarantees of the greedy algorithm.
An \emph{independence system} is a pair $(E, \mathcal{I})$ where $\mathcal{I} \subseteq 2^{E}$ is a collection of subsets of the ground set $E$ with the following properties (1) $\emptyset \in \mathcal{I}$ and (2) the collection $\mathcal{I}$ is downward closed, i.e. if $A \subseteq B$ and $B \in \mathcal{I}$, then $A \in \mathcal{I}$.
We remark that the edge sets of degree-bounded forests form such an independent system.

\begin{remark}
    \label{rmk:indepencen-system-BDF}
    Let $(G = (V, E), w, b)$  be an instance of $\DBFs$. Then, the pair $(E, \mathcal{F})$, where 
    \[
    \mathcal{F} \ \coloneqq\ \{ F \subseteq E \colon F \mbox{ acyclic  and } \deg_F(v) \leq b_v \ \text{ for all } v\in V \}
    \] is an independence system.
\end{remark}

Korte and Hausmann proved that the approximation factor of the greedy algorithm can be bounded in terms of the rank quotient of the independent system:

\begin{theorem}[Hausmann, Korte \cite{Korte-Hausmann:1978:Analysis}]
    \label{thm:greedyindependence}
    Let $(E,\mathcal{I})$ be an independence system, and let $w \colon E \to \mathbb{R}_{\geq 0}$ be a weight function. Let $F_g$ be the output of the greedy algorithm and let $F^\ast$ be a maximum-weight independent set. Then,
    \begin{equation*}
        \frac{w(F_g)}{w(F^\ast)}\ \geq\ \min_{X\subseteq E}\frac{{\rm lr}(X)}{{\rm ur}(X)}
    \end{equation*}
    where 
    \begin{align*}
        {\rm lr}(X) \ \coloneqq&\ \min \ \big\{|A|  \ : A \text{ inclusion-wise maximal independent set of }X\big\}  \\
        {\rm ur}(X) \ \coloneqq&\ \max \big\{|A| \ : A \text{ inclusion-wise maximal independent set of }X \big\}. 
    \end{align*}
\end{theorem}

An immediate corollary of \Cref{lem:greedyratio} and \Cref{thm:greedyindependence} is that the greedy algorithm has approximation factor at most $2$.
\begin{proof}[Proof of \Cref{thm:greedy-apx}]
    By \Cref{rmk:indepencen-system-BDF}, the degree-bounded forests correspond to an independent system.
    Consider a set $X\subseteq E$.
    By \Cref{lem:greedyratio} applied to the graph $(V,X)$, the ratio between the size of two inclusion-wise maximal forests in $(V,X)$ that satisfy the degree bounds is at most $2$. Thus, we have 
    $\frac{\textrm{lr}(X)}{\text{ur}(X)} \geq \frac{1}{2}$ and \Cref{thm:greedyindependence} concludes the proof.
\end{proof}

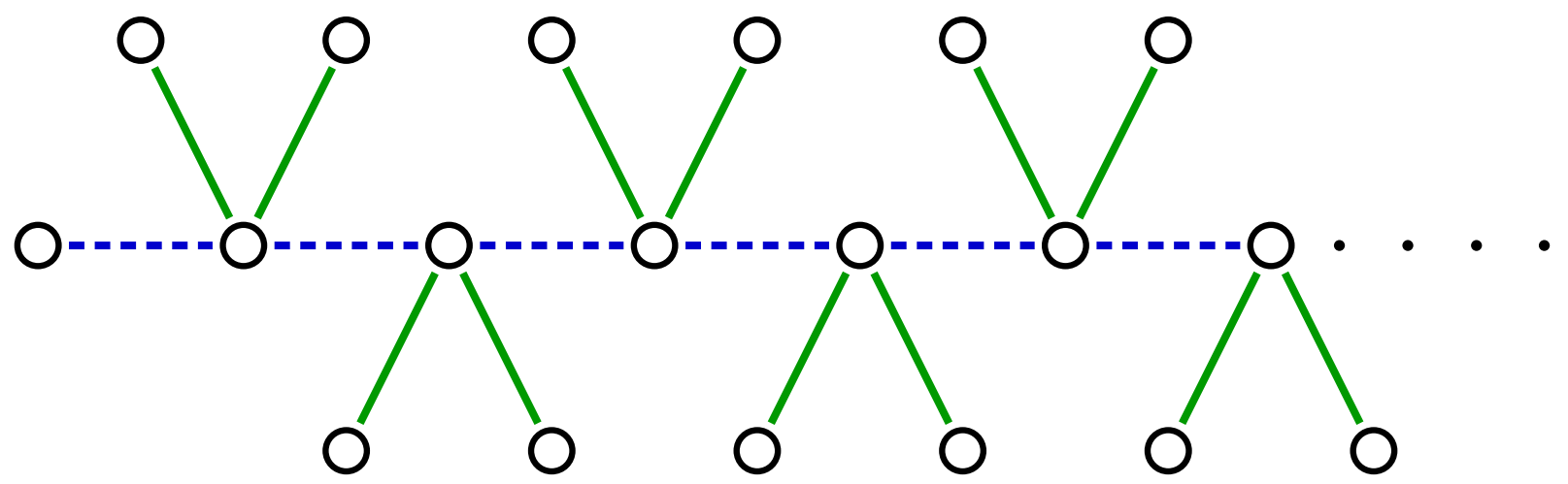
\begin{figure}[h!]
    \centering
    \begin{tikzpicture}[xscale=1.4,yscale=1.4,
two/.style={very thick,draw=black,circle,minimum size=8,inner sep=2pt, outer sep=2pt}
]

\node [two] (u1) at (0.5,1) {};
\node [two] (u2) at (1.5,1) {};
\node [two] (u3) at (2.5,1) {};
\node [two] (u4) at (3.5,1) {};
\node [two] (u5) at (4.5,1) {};
\node [two] (u6) at (5.5,1) {};

\node [two] (l1) at (1.5,-1) {};
\node [two] (l2) at (2.5,-1) {};
\node [two] (l3) at (3.5,-1) {};
\node [two] (l4) at (4.5,-1) {};
\node [two] (l5) at (5.5,-1) {};
\node [two] (l6) at (6.5,-1) {};

\node [two] (v1) at (0,0) {};
\node [two] (v2) at (1,0) {};
\node [two] (v3) at (2,0) {};
\node [two] (v4) at (3,0) {};
\node [two] (v5) at (4,0) {};
\node [two] (v6) at (5,0) {};
\node [two] (v7) at (6,0) {};

\begin{scope}[line width =1.5pt]
    \begin{scope}[blue!80!black, densely dashed]
        \draw (v1) to (v2);
        \draw (v2) to (v3);
        \draw (v3) to (v4);
        \draw (v4) to (v5);
        \draw (v5) to (v6);
        \draw (v6) to (v7);
    \end{scope}
    \begin{scope}[green!60!black]
        \draw (v2) to (u1);
        \draw (v2) to (u2);
        \draw (v4) to (u3);
        \draw (v4) to (u4);
        \draw (v6) to (u5);
        \draw (v6) to (u6);
        \draw (v3) to (l1);
        \draw (v3) to (l2);
        \draw (v5) to (l3);
        \draw (v5) to (l4);
        \draw (v7) to (l5);
        \draw (v7) to (l6);
    \end{scope}
\end{scope}

\node [circle, fill, inner sep=0.75pt] at (6.333,0) {};
\node [circle, fill, inner sep=0.75pt] at (6.666,0) {};
\node [circle, fill, inner sep=0.75pt] at (7,0) {};
\node [circle, fill, inner sep=0.75pt] at (7.33,0) {};

\end{tikzpicture}
    \vspace{0.5cm}
    \caption{A family of graphs showing that the approximation factor of $2$ for the greedy algorithm is tight, even for unit-weight graphs. Each vertex $v$ has a degree bound of $b_v = 2$ and each edge has cost $1$. The greedy algorithm might select the blue dotted edges whereas an optimal solution consists of the green (solid) edges.}
    \label{fig:tightgreedy}
\end{figure}

The approximation guarantee of the greedy algorithm from \Cref{thm:greedy-apx} is tight, as the example in \Cref{fig:tightgreedy} shows. 

\newpage
\section{Improving on Greedy}\label{sec:improving_on_greedy}
    In this section, we provide an $\sfrac{8}{5}$-approximation algorithm for the maximum-weight degree-bounded forest problem. The approximation ratio improves to $\min\{ \frac{\Delta +1}{2}, \frac{\Delta + 1}{\Delta}\}$ when $b_v = \Delta$ for all $v$. We first present our algorithm in \Cref{sec:algo}. In \Cref{sec:degree_violation,sec:b_matching}, we analyze the two main subroutines of our algorithm, which we then use to determine the overall approximation factor in \Cref{sec:apx_factor}.
    Finally, in \Cref{sec:integrality_gap}, we derive an upper bound on the integrality gap of the natural linear programming relaxation with blossom constraints.
    
    \subsection{Algorithm}\label{sec:algo}
    First, we observe that we may assume that the given graph $G$ is simple as we can otherwise delete all but the maximum-weight edge from every set of parallel edges
    To improve upon the greedy algorithm, we use two complementary approaches. First, we leverage Lau and Singh's iterative method to produce a forest $T'$  that violates degree bounds by an additive $+1$ term.
    More precisely, their algorithm provides the following guarantee, where we write $F^\ast$ to denote a fixed optimal solution to the given $\DBFs$ instance.
    \begin{theorem}[Adaptation of Theorem 1.2 \cite{Singh-Lau:2015:Approximating}]
        \label{thm:LPapprox}
        Given a graph $G=(V,E)$ with edge weights $w\colon E \to \mathbb{R}_{\geq 0}$ and degree bounds $b \colon V \to \mathbb{N}$, we can in polynomial time compute a forest $F$ such that 
        \begin{itemize}
            \item $w(F) \geq w(F^\ast)$, and
            \item  for each vertex $v \in V$, we have $\deg_{F}(v) \leq b_v + 1$.
        \end{itemize}
    \end{theorem}
    \begin{proof}
        [Proof of \Cref{thm:LPapprox}]
        The result of \cite{Singh-Lau:2015:Approximating} applies to the minimum-cost degree-bounded spanning tree problem. Their algorithm outputs a minimum-cost spanning tree with additive degree violation at most $1$, whenever there exists a feasible solution to the problem (Theorem 1.2 in  \cite{Singh-Lau:2015:Approximating}).
        As they do not require the edge costs to be nonnegative, their result also applies to the maximization version of the problem (by negating the costs).
        It remains to reduce the claimed statement for the maximum-weight degree-bounded forest problem to the analogous statement for the maximum-weight degree-bounded spanning tree problem.
        
        Given an instance $(G = (V, E), w, b)$ of $\DBFs$, we modify it by adding a new vertex $v'$ adjacent to $v$ for each $v \in V$.
        Let the weight of the new edge $\{v, v'\}$ be a sufficiently large constant $M$, for instance $M > w(E)$. 
        We add a root $r$ adjacent to each newly added vertex $v'$, where each edge $\{r, v'\}$ has weight $0$. Finally, we define $b'_v \leftarrow b_v + 1$ for all $v\in V$ and let all degree bounds $b'_v$ for newly added vertices be infinite (or equal to the number of vertices of the new graph).
        
        Let $G' = (V',E')$ denote the new graph. 
        Observe that the optimal degree-bounded forest $F^* \subseteq E$ in $G$ can be extended into a tree $T^*$ in $G'$ whose weight is equal to $w(T^*) = w(F^*) + |V|\cdot M$ and satisfies the degree-bounds $b'$.
        Moreover, because $M$ is sufficiently large, any spanning tree $T$ with $w(T) \geq w(T^*)$ includes all edges $\{v, v'\}$ between vertices $v\in V$ and their newly added copy $v'\in V'\setminus V$. 
        In particular, this applies to the output $T$ of Lau-Singh's algorithm on $G'$ with degree bounds $b'$.
        Thus, restricting the tree $T$ to $G$ yields a forest $F$ satisfying $w(F) \geq w(F^*)$ and $\deg_{F}(v) \leq b_v + 1$ for all $v\in V$.
    \end{proof}
    We refer to the algorithm from \Cref{thm:LPapprox} as \textsc{Lau--Singh}, and we will apply it with some auxiliary weight function $w'$.
    We then prune some edges by computing a maximum-weight subset of $F$ respecting the degree bounds (with respect to the original weight $w$). 
    This approach improves upon greedy when degree bounds are high.
    The complementary approach is to compute a maximum weight $b$-matching $M$ with respect to another auxiliary weight function $w''$.
    \begin{definition}
    [$b$-matching]
    \label{def:b-matching}
    Given a graph $G = (V, E)$ and degree bounds $b \colon V \to \mathbb{N}$, a $b$-matching $M \subseteq E$ is a set of edges such that $\deg_M(v) \leq b_v$ for all $v\in V$.
    \end{definition}

    \begin{theorem}
    [\cite{Padberg-Rao:1982:Odd}]
    \label{thm:b-matching}
        There is a polynomial time algorithm to compute a maximum-weight $b$-matching in a given weighted graph with degree bounds $b$.
    \end{theorem}   
    Again, we will then prune edges by computing a maximum-weight forest $F'' \subseteq M$. The pseudocode of our algorithm is given in \Cref{alg:nonuniformb}.
    
    To define the auxiliary weight functions $w'$ and $w''$, we introduce the following parameter.
    \begin{definition}
       Given a graph $G = (V, E)$ and degree bounds $b \colon V \to \mathbb{N}$, we define $\bmin(e) \coloneqq \min \{ b_u, b_v \}$ for any edge $e = \{u, v\} \in E$.
    \end{definition}

Intuitively, by working with the auxiliary weight functions, we give preference not only to those edges with high original weight, but also try to choose an edge set that requires little pruning in the second step.
In particular, in the Lau-Singh approach we favor edges with large $b_{\min}$, whereas in the $b$-matching approach we favor edges with small $b_{\min}$.
The reason for the exact choice of the weight functions will become clear from the analysis of the algorithm.
    
    
    \begin{algorithm}[h!]
    \KwIn{A simple weighted graph $G=(V,E,w)$ and degree bounds $b\colon V\to\mathbb{N}$.}
    \KwOut{A forest $\bar F\subseteq E$ satisfying the degree bounds $b$.}
    
    \ForEach{$e\in E$}{
        $w'_e \gets \frac{\bmin(e)}{\bmin(e)+1} \cdot w_e$,\; \mbox{ and }
        $w''_e \gets \frac{1}{\bmin(e)} \cdot w_e$.\;
    }
    $T' \gets \textsc{Lau--Singh}(G,w')$ \mbox{ and }
    $F' \gets$ an optimal solution to $\DBFs$ on $(V,T',w)$\; \label{line:1}

    $T'' \gets \textsc{Maximum-Weight-$b$-Matching}(G,w'')$\; \mbox{ and } 
    $F'' \gets$ an optimal solution to $\DBFs$ on $(V,T'',w)$\; \label{line:2}
    \Return $\arg\max \{w(F'), w(F'') \}$
    \caption{Approximation algorithm for $\DBF$}
    \label{alg:nonuniformb}
\end{algorithm}

\begin{proposition}
        \label{prop:runtime}
        \Cref{alg:nonuniformb} runs in polynomial time. 
    \end{proposition}
    \begin{proof}[Proof of \Cref{prop:runtime}]
        \textsc{Lau--Singh} runs in polynomial time (\Cref{thm:LPapprox}). Computing a maximum weight $b$-matching can be solved exactly in polynomial time (\Cref{thm:b-matching}).
        Let $T', T''$ be the output of the respective algorithms.
        Because $T'$ is acyclic, the problem of finding an optimal solution in $(V, T', w)$ reduces to finding a maximum-weight $b$-matching (in a tree).
        Similarly, finding an optimal solution to $\DBFs$ on $(V, T'', w)$ is equivalent to computing a maximum-weight forest on $(V, T'', w)$ because $(V, T'')$ satisfies the degree bounds.
    \end{proof}

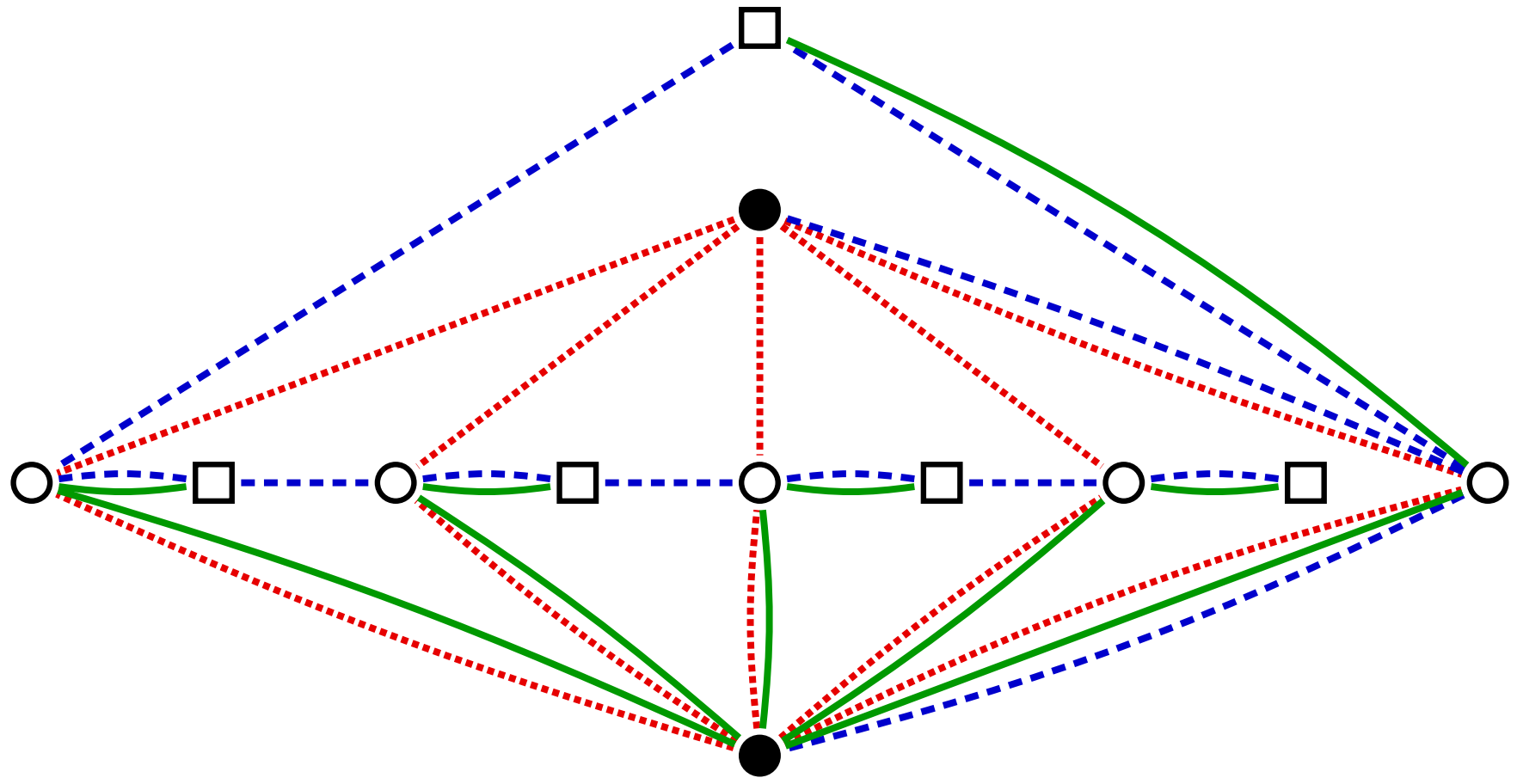
\begin{figure}[h!]
        \centering
        \begin{tikzpicture}[xscale=1.4,yscale=1.4,
one/.style={very thick,draw=black,minimum size=8,inner sep=2pt, outer sep=2pt},
many/.style={very thick,draw=black,fill=black,circle,minimum size=8,inner sep=2pt, outer sep=2pt},
two/.style={very thick,draw=black,circle,minimum size=8,inner sep=2pt, outer sep=2pt},
]

\node[two] (v1) at (0,0) {};
\node[one] (v2) at (1,0) {};
\node[two] (v3) at (2,0) {};
\node[one] (v4) at (3,0) {};
\node[two] (v5) at (4,0) {};
\node[one] (v6) at (5,0) {};
\node[two] (v7) at (6,0) {};
\node[one] (v8) at (7,0) {};
\node[two] (v9) at (8,0) {};
\node[one] (v10) at (4,2.5) {};

\node[many] (w1) at (4,1.5) {};
\node[many] (w2) at (4,-1.5) {};

\begin{scope}[line width =1.5pt]
    \begin{scope}[red!90!black, densely dotted]
    \draw (w1) to (v1);
    \draw [bend left=4] (w2) to (v1);
    \draw (w1) -- (v3);
    \draw [bend left=4] (w2) to (v3);
    \draw (w1) -- (v5);
    \draw [bend left=6] (w2) to (v5);
    \draw (w1) -- (v7);
    \draw [bend left=4] (w2) to (v7);
    \draw[bend right=3] (w1) to (v9);
    \draw[bend left=6] (w2) to (v9);
    \end{scope}
    \begin{scope}[blue!80!black, densely dashed]
    \draw[bend left=8] (v1) to (v2);
    \draw (v2) -- (v3);
    \draw[bend left=8] (v3) to (v4);
    \draw (v4) -- (v5);
    \draw[bend left=8] (v5) to (v6);
    \draw (v6) -- (v7);
    \draw[bend left=8] (v7) to (v8);
    \draw (v9) -- (v10);
    \draw (v10) -- (v1);
    
    \draw[bend right=3] (v9) to (w1);
    \draw[bend left=6] (v9) to (w2);
    \end{scope}
    \begin{scope}[green!60!black]
    \draw [bend right=4] (w2) to (v1);
    \draw [bend right=4] (w2) to (v3);
    \draw [bend right=6] (w2) to (v5);
    \draw [bend right=4] (w2) to (v7);
    \draw (w2) -- (v9);
    \draw[bend right=8] (v9) to (v10);
    
    \draw[bend right=8] (v1) to (v2);
    \draw[bend right=8] (v3) to (v4);
    \draw[bend right=8] (v5) to (v6);
    \draw[bend right=8] (v7) to (v8);
    \end{scope}
\end{scope}
    
\end{tikzpicture}
        \vspace{0.5cm}
        \caption{
        \label{fig:weight_functions}
        Illustration of a family of examples for which the approximation factor of our algorithm would become arbitrarily close to $2$ if we worked with the original weight function $w$ instead of $w'$ and $w''$.
        Here every edge $e$ has weight $w_e=1$. 
        The $k$ vertices shown as white squares have a degree bound of $1$, the $k$ vertices shown as white circles have a degree bound of $2$, and the two vertices shown as filled black circles have an irrelevant degree bound of $|V|$.
        An optimal solution with weight $2k=10$ is shown with green (solid) edges.
        A maximum-weight $b$-matching with weight $2k$ is shown in red (dotted). A maximum-weight forest within this $b$-matching has weight $k+1$.
        A forest $F$ satisfying the conditions of \Cref{thm:LPapprox} is shown in blue (dashed). 
        Every subset of this forest that satisfies the degree bounds has weight at most $k+1$.
        }
    \end{figure}

We highlight that the use of the auxiliary weight functions $w'$ and $w''$ is crucial.
    As the example in \Cref{fig:weight_functions} shows, the algorithm might otherwise not be better than a $2$-approximation.

\subsection{Removing degree bound violations}\label{sec:degree_violation}
    In this section, we derive a lower bound for the weight $w(F')$ of the solution obtained via pruning the result $T'$ of the algorithm by Lau and Singh. 
    More precisely, we show how this weight relates to that of the optimal solution $F^\ast$.
    \begin{lemma}
        \label{lem:w1}
        The solution $F'$ computed in Line~\ref{line:1} of \Cref{alg:nonuniformb}  satisfies
        \begin{align*}
            w(F') & \ \geq\ w'(F^\ast) \ =\ \sum_{e \in F^\ast} \frac{\bmin(e)}{\bmin(e) + 1} \cdot w_e.
        \end{align*}
    \end{lemma}
    As a consequence of \Cref{lem:w1}, we obtain an improved approximation factor whenever the minimum degree bound is at least $2$.
    \begin{corollary}
        \label{cor:w1-unif}
        Let $\Delta \geq 1$ such that $b_v \geq \Delta$ for all $v\in V$. 
        Then the solution  $F'$ computed in Line~\ref{line:1} of \Cref{alg:nonuniformb} satisfies
        \begin{align*}
            w(F) & \geq \frac{\Delta}{\Delta +1} \cdot w(F^\ast).
        \end{align*}
    \end{corollary}
    \begin{remark}
        This implies a $\frac{\Delta + 1}{\Delta}$-approximation algorithm for uniform $\DBFs$ with degree $\Delta$. Additionally, it yields an alternative $2$-approximation for $\DBFs$.
    \end{remark}

In order to prove \Cref{lem:w1}, we need to show that there exists a set $F \subseteq T'$ satisfying the degree bounds and $w(F) \geq w'(F^\ast)$.
To this end, we will use a probabilistic argument.
We will construct a distribution of sets $F \subseteq T'$ satisfying the degree bounds with expected weight at least $w'(F^\ast)$.

\begin{proof}[Proof of \Cref{lem:w1}]
    Recall that $w'_e \coloneqq \frac{\bmin(e)}{\bmin(e) + 1} \cdot w_e$ for all $e \in E$. 
    By \Cref{thm:LPapprox}, the forest $T'$ satisfies $\deg_{T'}(v) \leq b_v + 1$ and $w'(T') \geq w'(F^\ast)$. 
    We will show that there exists a probability distribution $\mathcal{D}$ over the set 
    \[
    \mathcal{F} = \{F \subseteq T' \mid \deg_F(v) \leq b_v \text{ for all } v \in V\}
    \]
    of feasible solutions contained in $T'$ such that for every edge $e\in T'$, we have
    \begin{equation}\label{eq:probability}
    \mathbb{P}_{F \sim \mathcal{D}}[e \in F] \ \geq\ \frac{b_{\min}(e)}{b_{\min}(e) + 1}.
    \end{equation}
    This will complete the proof of \Cref{lem:w1} because $w(F') \geq \mathbb{E}_{F \sim \mathcal{D}}[w(F)] \geq w'(T') \geq w'(F^\ast)$.
    
    It remains to show that such a distribution $\mathcal{D}$ of forests exists.
    Choose $M\in\mathbb N$ such that $b_v+1$ divides $M$ for every $v\in V$. We will construct $M$ forests
    $F_1,\ldots,F_M\subseteq T'$ satisfying the degree bounds, while every edge $e\in F'$ belongs to at least $\frac{\bmin(e)}{\bmin(e)+1}\cdot M$ of these forests. The desired distribution is then the uniform distribution over
    $\{F_1,\ldots,F_M\}$.
    
    For each edge $e\in T'$, we define a deletion set $I(e)\subseteq[M]$; the forest $F_i$ consists of all edges whose deletion set does not contain $i$. Formally, $F_i \coloneqq \{e\in T'\mid i\notin I(e)\}$.
    To construct the deletion sets, we will choose for every vertex $v$ violating its degree bound and every index $i\in [M]$, an incident edge $e\in \delta_{T'}(v)$ to be deleted. 
    This ensures that the resulting forest $F_i$ satisfies the degree bounds.
    Our choice will be encoded in the local deletion sets $I_v(e)$ that we will construct for every vertex $v$ and every incident edge $e\in\delta_{T'}(v)$.
    We then set
    \begin{equation*}
    I(e)=I_u(e)\cup I_v(e), \qquad \mbox{ for all } e=\{u,v\} \in T'.
    \end{equation*}
    In particular, if $i \in I_v(e)$, then the edge $e$ is not contained in the forest $F_i$.

    To construct the local deletion sets, we choose an arbitrary root of every connected component of the forest $T'$, and process each such tree from the root to the leaves.
    For a vertex $v$, if $\deg_{T'}(v)\le b_v$, we simply set $I_v(e)=\emptyset$ for every edge $e \in \delta_{T'}(v)$.
    
    Now suppose $\deg_{T'}(v)=b_v+1$. 
    Then we partition $[M]$ into $b_v + 1$ sets, each of size $s_v \coloneqq \frac{M}{b_v+1}$.
    If $v$ is the root, we simply choose an arbitrary partition of $[M]$ into $b_v+1$ sets of size $s_v$ and assign one set to each of the $b_v+1$ incident edges $e\in \delta_{T'}(v)$ to obtain the sets $I_v(e)$.
    Otherwise, let $u$ be the parent of $v$ and let $e^\ast=\{u,v\}$. 
    Then the set $I_u(e^\ast)$ has already been constructed. 
    We choose the set $I_v(e^\ast)$ assigned to the edge $e^\ast$ so that it is nested with $I_u(e^\ast)$. Because we want to have $|I_v(e^\ast)| = s_v$, this means we ensure
    \begin{equation*}
    I_u(e^\ast)\subseteq I_v(e^\ast) \quad\text{if }|I_u(e^\ast)|\le s_v,
    \end{equation*}
    and
    \begin{equation*}
    I_v(e^\ast)\subseteq I_u(e^\ast) \quad\text{otherwise}.
    \end{equation*}
    Observe that such a choice of $I_v(e^\ast)$ is always possible by either enlarging $I_u(e^\ast)$ to size $s_v$ or selecting a
    subset of $I_u(e^\ast)$ of size $s_v$. 
    Then we partition the remaining elements of $[M]$ arbitrarily into $b_v$ additional sets $I_v(e)$ of size $s_v$ each, one for each remaining incident edge $e\in \delta_{T'}(v) \setminus \{e^\ast\}$.

    It is straightforward to prove that each forest $F_i$ is feasible, i.e.\ satisfies $\deg_{F_i}(v) \leq b_v$. Indeed, for any vertex $v \in V$ with $\deg_{T'}(v) = b_v+1$, the sets $I_v(e)$ associated with $e \in \delta_{T'}(v)$ partition $[M]$. Therefore, for every $i \in [M]$ at least one edge incident to $v$ will be deleted.
    Next, we compute the probability of an edge $e = \{u, v\} \in T'$ being deleted. Since $F_i$ is sampled uniformly, this probability is equal to the probability of sampling a random index of $[M]$ contained in $I(e)$. Since the sets $I_u(e)$ and $I_v(e)$ are nested, we have
    \begin{align*}
        \mathbb{P}[e \notin F] & = \frac{|I(e)|}{M} = \frac{|I_u(e)\cup I_v(e)|}{M} \ = \ \frac{\max\big\{|I_u(e)|,|I_v(e)|\big\}}{M} \ \leq\ \frac{\frac{M}{\bmin(e) + 1}}{M} = \frac{1}{\bmin(e) + 1}. \qedhere
    \end{align*}
\end{proof}

\subsection{Removing cycles}\label{sec:b_matching}
In this section, we provide a lower bound on the weight of the solution $F''$ obtained from the $b$-matching $T''$.
Recall that $F^\ast$ denotes a fixed optimal solution to the given $\DBFs$ instance.
Similar to the proof of \Cref{lem:w1}, we will again use a probabilistic argument.

\begin{lemma}
    \label{lem:w2}
   The solution $F''$ computed in Line~\ref{line:2} of \Cref{alg:nonuniformb}  satisfies
    \begin{align*}
        w(F'') & \ \geq\  w''(F^\ast) \ =\ \sum_{e \in F^\ast} \frac{1}{\bmin(e)} \cdot w_e.
    \end{align*}
\end{lemma}
\begin{proof}[Proof of \Cref{lem:w2}]
    Recall that $T''$ is a maximum weight $b$-matching with respect to $w''$, where $w''_e \coloneqq \frac{1}{\bmin(e)} \cdot w_e$ for all $e \in E$. 
    In particular, we have  $w''(T'') \geq w''(F^\ast)$.
    We show that there is a probability distribution $\mathcal{D}$ over 
    \[
    \mathcal{F} \ =\ \{F \subseteq T'' \mid F \mbox{ forest and }\deg_F(v) \leq b_v \text{ for all } v \in V\}
    \]
    such that for each edge $e \in T''$, we have
    \begin{equation}\label{eq:probability2}
    \mathbb{P}_{F \sim \mathcal{D}}[e \in F] \ \geq\ \frac{1}{b_{\min}(e)}.
    \end{equation}
    By \Cref{alg:nonuniformb}, the edge set  $F'' \subseteq T''$ is a maximum-weight forest in $T''$. 
    Thus, we have 
    \begin{align*}
        w(F'') \ \geq\ \mathbb{E}_{F\sim \mathcal{D}} \big[ w(F) \big] \ \geq\ \sum_{e \in T''} \frac{1}{b_{\min}(e)}\, w_e \ =\ w''(T'') \ \geq\ w''(F^\ast).
    \end{align*}

    It remains to prove the existence of a probability distribution $\mathcal{D}$ with the properties claimed above.
    We construct $\mathcal{D}$ by defining a randomized procedure to sample a forest $F \subseteq T''$.
    First, fix an arbitrary total order $\prec$ on the vertices $V$. 
    For each edge $e = \{u,v\} \in T''$, orient $e$ toward the endpoint with the smaller degree bound, breaking ties consistently via the order $\prec$. In particular, orient $e$ towards $u$ if $b_u < b_v$ or if $b_u = b_v$ and $u \prec v$.  Otherwise, orient $e$ towards $v$.
    Along an edge $e$, the set $\{(b_v, \prec_v)\}_{v \in e}$ is then lexicographically ordered, where $\prec_v$ is the position of the vertex $v$ with respect to the order $\prec$.
    Observe that there is no directed cycle after the orientation. Indeed, a directed cycle $C$ implies the existence of a strictly increasing sequence that loops back to itself, leading to a contradiction.
    For each vertex $v\in V$, let $\delta^{-}(v)$ be the edges in-going into $v$, and note that $|\delta^{-}(v)| \leq \bmin(e)$ for every edge $e\in \delta^-(v)$ by the choice of the orientation.
    The distribution $\mathcal{D}$ over degree-bounded forests is then obtained by sampling for each $v \in V$ with $\delta^-(v)\neq \emptyset$, an edge $e \in \delta^{-}(v)$ uniformly at random (and omitting the orientation). 
    Because we sample only a single incoming edge for each vertex and no directed cycle exists, the resulting edge set $F$ is a forest.
    Moreover, we get that for each edge $e$ oriented towards a vertex $v$,
    \begin{align*}
        \mathbb{P}_{F \sim \mathcal{D}}\big[e \in F\big] &\ =\ \frac{1}{|\delta^{-}(v)|}\ \geq\ \frac{1}{b_{\min}(e)}. \qedhere
    \end{align*}
\end{proof}

For uniform $\DBFs$, that is, when there is some $\Delta \geq 1$ such that $b_v=\Delta$ for all $v\in V$, we can improve \Cref{lem:w2}.
To this end, we build on the Nash--Williams theorem for partitioning a graph into edge-disjoint forests. 
Here we state a direct consequence of this theorem for covering a graph fractionally with forests.
\begin{theorem}[Corollary 5.4.3 of \cite{Scheinerman-Ullman:2011:Fractional}]
    \label{fractonaltreepacking}
    Let $G = (V,E)$ be a graph, and let $k = \max_{H\subseteq G}\frac{|E(H)|}{|V(H)|-1}$.
    Then $G$ can be fractionally covered by $k$ forests. 
    More precisely, there is a collection $F_1,\dots,F_\ell$ of forests and nonnegative weights $\lambda_1,\dots,\lambda_\ell$ such that $\sum_{i=1}^\ell \lambda_i = k$ and $\sum_{i:e\in F_i}\lambda_i \geq1$ for all $e\in E$.
\end{theorem}

Using \Cref{fractonaltreepacking}, we can now derive our improved lower bound on $w(F'')$ for the uniform special case of $\DBFs$.
Observe that in this special case, we have $w'' = \frac{1}{\Delta} \cdot w$ and hence we could equivalently replace $w''$ by the original weight function $w$ in \Cref{alg:nonuniformb}.

\begin{lemma}
    \label{lem:unif-w2}    
    Let $\Delta \geq 1$ and suppose that $b_v = \Delta$ for all $v \in V$ and some $\Delta \geq 1$.
    Then the solution $F''$ computed in Line~\ref{line:2}  of \Cref{alg:nonuniformb}  satisfies
    \begin{align*}
        w(F'') & \geq \frac{2}{\Delta + 1} \cdot w(F^\ast).
    \end{align*}
\end{lemma}
\begin{proof}
    [Proof of \Cref{lem:unif-w2}]
    Observe that $T''$ is a maximum-weight $b$-matching with respect to $w''$ and hence in particular $w(T'') \geq w(F^\ast)$.
    We apply \Cref{fractonaltreepacking} to obtain forests $F_1,\dots,F_{\ell}$ and non-negative weights $\lambda_1,\dots,\lambda_{\ell}$.
    The maximum-weight forest $F'' \subseteq T''$ satisfies $w(F'') \geq w(F_i)$ for all $i \in [\ell]$.
    We now show that $w(F'') \geq \frac{1}{k}\, w(F^\ast)$, where $k = \max_{H\subseteq M}\frac{|E(H)|}{|V(H)|-1}$. 
    Indeed, we have
    \begin{align*}
    w(F^\ast) & \ \leq\ w(T'') \ =\ \sum_{e\in T''} w_e \ \leq\ \sum_{e\in E}w_e \cdot  \sum_{i:e\in F_i}\lambda_i \ =\ \sum_{i=1}^\ell \lambda_i \cdot w(F_i) \ \leq\ \sum_{i=1}^\ell \lambda_i \cdot w(F'') \ =\ k \cdot w(F'').
    \end{align*}
    To complete the proof, we will show $k \leq \frac{\Delta + 1}{2}$. 
    Consider any subgraph $H$ of the forest $T''$.
    We need to show $|E(H)| \leq \frac{\Delta + 1}{2} \cdot (|V(H)| -1)$.
    To this end, we distinguish two cases.\\
    
    \noindent \textbf{Case 1: $|V(H)|\leq \Delta +1$}\\
    Because the graph given in our $\DBFs$ instance is simple, the same applies to the subgraph  $H$.
    Thus,
        \begin{align*}
        |E(H)| \ \leq\  |V(H)| \cdot \frac{|V(H)|-1}{2} \ \leq\ \frac{\Delta + 1}{2} \cdot \big(|V(H)|-1\big).
        \end{align*}
        
    \noindent \textbf{Case 2: $|V(H)|> \Delta +1$}\\
    Every vertex $v$ of $H$ satisfies $\deg_H(v) \leq \deg_{T''}(v) \leq \Delta$. 
    This implies
    \begin{align*}
    |E(H)| \ \leq\ \frac{|V(H)|\cdot\Delta}{2}\ \leq\ \frac{\Delta + 1}{2}\cdot (|V(H)|-1).
    \end{align*}
    where the last inequality follows from $|V(H)|> \Delta +1$.
\end{proof}

\subsection{Determining the approximation factor}\label{sec:apx_factor}

Finally, we combine \Cref{lem:w1} and \Cref{lem:w2} to determine the approximation factor of \Cref{alg:nonuniformb}.
\begin{theorem}
    \label{thm:5-8}
    \Cref{alg:nonuniformb} is an $\frac{8}{5}$-approximation algorithm for $\DBFs$.
\end{theorem}
\begin{proof}
Recall that $F^\ast$ denotes a fixed optimal solution to $\DBFs$.
By \Cref{lem:w1} and \Cref{lem:w2}, we have
\begin{align*}
\max \Big\{ w(F'),\, w(F'') \Big\} \ \geq&\ \frac{3}{4}\cdot w(F') + \frac{1}{4}\cdot w(F'')  \\
\ \geq &\ \sum_{e\in F^*} \left(\frac{3}{4} \cdot \frac{b_{\min}(e)}{b_{\min}(e) +1} + \frac{1}{4} \cdot \frac{1}{b_{\min}(e)} \right) \cdot w_e \\
\geq&\ \sum_{e\in F^*} \frac{5}{8} \cdot w_e,
\end{align*}
where the last inequality uses that for every $k\in \mathbb{N}$, we have $f(k) \coloneqq \frac{3}{4} \cdot \frac{k}{k+1} + \frac{1}{4} \cdot \frac{1}{k} \geq \frac{5}{8}$.
To see this, observe that $f(1)=f(2)= \frac{5}{8}$ and $f(k)$ is monotonically increasing for $k \geq 2$.
\end{proof}

\begin{figure}[h!]
        \centering
        \begin{tikzpicture}[xscale=1.4,yscale=1.4,
one/.style={very thick,draw=black,minimum size=8,inner sep=2pt, outer sep=2pt},
many/.style={very thick,draw=black,fill=black,circle,minimum size=8,inner sep=2pt, outer sep=2pt},
two/.style={very thick,draw=black,circle,minimum size=8,inner sep=2pt, outer sep=2pt},
]

\node[many] (v1) at (0,0) {};
\node[two] (v2) at (1,0) {};
\node[two] (v3) at (2,0) {};
\node[two] (v4) at (3,0) {};
\node[two] (v5) at (4,0) {};
\node[two] (v6) at (5,0) {};
\node[many] (v7) at (6,0) {};

\node[one] (w1) at (7, -1) {};
\node[one] (t1) at (8, -0.5) {};
\node[one] (t2) at (8, -1.5) {};

\node[one] (w2) at (7, 1) {};
\node[one] (t3) at (8, 1.5) {};
\node[one] (t4) at (8, 0.5) {};

\begin{scope}[line width =1.5pt]
    \begin{scope}[red!90!black, densely dotted]
    \draw[bend left=20] (v1) to (v2);
    \draw[bend left=30] (v1) to node [sloped, pos=0.8, above] {$1$} (v3);
    \draw[bend left=40] (v1) to node [sloped, pos=0.7, above] {$1$} (v4);
    \draw[bend left=60] (v1) to node [sloped, pos=0.58, above] {$1$} (v5);
    \draw[bend left=80] (v1) to node [sloped, pos=0.5, above] {$1$} (v6) ;

    \draw[bend left=80] (v7) to (v2);
    \draw[bend left=60] (v7) to (v3);
    \draw[bend left=40] (v7) to (v4);
    \draw[bend left=20] (v7) to (v5);
    \draw[bend left=10] (v7) to (v6);

    \draw[bend left=10] (t1) to node [midway, right] {$2$} (t2);
    \draw[bend left=10] (t3) to node [midway, right] {$2$} (t4);
    
    \end{scope}
    
    \begin{scope}[blue!80!black, densely dashed]
    \draw[bend left=10] (v1) to (v2);
    \draw[bend left=10] (v2) to (v3);
    \draw[bend left=20] (v2) to (v4);
    \draw[bend left=20] (v4) to node [midway, sloped, above] {$2$} (v6);
    \draw[bend left=10] (v5) to (v6);
    \draw[bend left=20] (v6) to (v7);

    \draw (w1) -- (t1) node [midway, sloped, above] {$1$};
    \draw (w1) to (t2);
    \draw (w2) -- (t3) node [midway, sloped, above] {$1$};
    \draw (w2) to (t4);
    
    \end{scope}
    
    \begin{scope}[green!60!black]
    \draw (v1) to node [midway, sloped, below] {$2$} (v2);
    \draw (v2) to node [midway, sloped, below] {$2$} (v3);
    \draw (v3) to node [midway, sloped, below] {$2$} (v4);
    \draw (v4) to node [pos=0.7,below] {$2$} (v5);
    \draw (v5) to (v6);
    \draw (v6) to (v7);

    \draw (t1) to node [midway, left] {$2$} (t2);
    \draw (t3) to (t4);
    \end{scope}
\end{scope}
    
\end{tikzpicture}
        \vspace{0.5cm}
        \caption{
        \label{fig:lp-tight}
        Illustration of a family of examples for which the approximation factor of our algorithm is tight (using \Cref{thm:LPapprox} as a black-box).
        In this example, all white squares vertices have a degree bound of $1$, all white circles have a degree bound of $2$, and all black vertices have an (irrelevant) degree bound of $|V|$.
        Because each edge $e$ in the left connected component satisfies $b_{\min}(e) =2$ and every edge in the other connected components (the triangles) satisfy $b_{\min}(e)=1$, for this particular family of instances, it does not matter whether we use the auxiliary weight functions $w'$ and $w''$ in the algorithm, or work only with $w$.
        \newline
        An optimal solution is composed of the solid green edges. For some odd $k\geq 3$ (here $k=3$), it consists of a path $P$ on $2^k - 1 = 7$ vertices (left) and $\lfloor\frac{1}{3}(2^k - 2)\rfloor = 2$ additional disjoint edges (right).
        Each green edge has a weight of $2$. In particular, the cost of the optimal solution is $w(F^\ast) = 2(2^k - 2) + 2 \cdot \lfloor\frac{1}{3}(2^k - 2)\rfloor \approx \frac{4}{3} \cdot 2^{k+1}$. In the figure, it has weight $16$.
        The output $T'$ of Lau-Singh's algorithm could be composed of dashed blue edges. On the vertices of the green path $P$, the blue edges form a binary tree whose root is the middle vertex of the path, where the weight of each edge of the binary tree is $2$.
        The blue edges of the triangles in the right have weight $1$ each. Observe that this blue forest violates the degree bounds by no more than an additive $1$.
        We have that $w(T') \geq w(F^\ast)$. 
        The forest $F'$ resulting from pruning the blue forest $T'$ has weight at most $\frac{5}{6}\cdot 2^{k+1}$ (in the figure, it has weight $10$). Indeed, let $T'_1, T_2'$ be the forest incident to squares (degree bound $1$), and circles (degree bound $2$), respectively. Let $F_1' \subseteq T_1', F_2' \subseteq T_2'$ be the forests after pruning. 
        Then, $w(F_1') = \sfrac{1}{2}\cdot w(T_1') = \sfrac{1}{2} \cdot 2 \cdot \lfloor\frac{1}{3}(2^k - 2)\rfloor$.
        Moreover, we have $w(F_2') = \frac{2}{3} \cdot 2^{k+1}$.
        \newline
        Finally, the output of the $b$-matching algorithm $T''$ could be composed of the dotted red edges. The edges have weight $1$ on the left part of the figure, except for the two outer edges that are also contained in the green optimal solution, and they have weight $2$ on the triangles (right). 
        One can verify that this is a maximum-weight $b$-matching of weight $2 \lfloor\frac{1}{3}(2^k - 2)\rfloor + (2^{k+1} - 2)$.
        A maximum-weight forest consisting of (dotted) red edges incident to circle vertices consists of two edges of weight $2$, and $2^{k} - 4$ edges of weight $1$.
        Thus, a maximum-weight degree-bounded red forest has weight $2^k + 2 \cdot \lfloor\frac{1}{3}(2^k - 2)\rfloor \approx \frac{5}{6} 2^{k+1}$. In the case $k=3$ shown in the figure, the max-weight degree-bounded red forest has weight $12$.
        }
\end{figure}
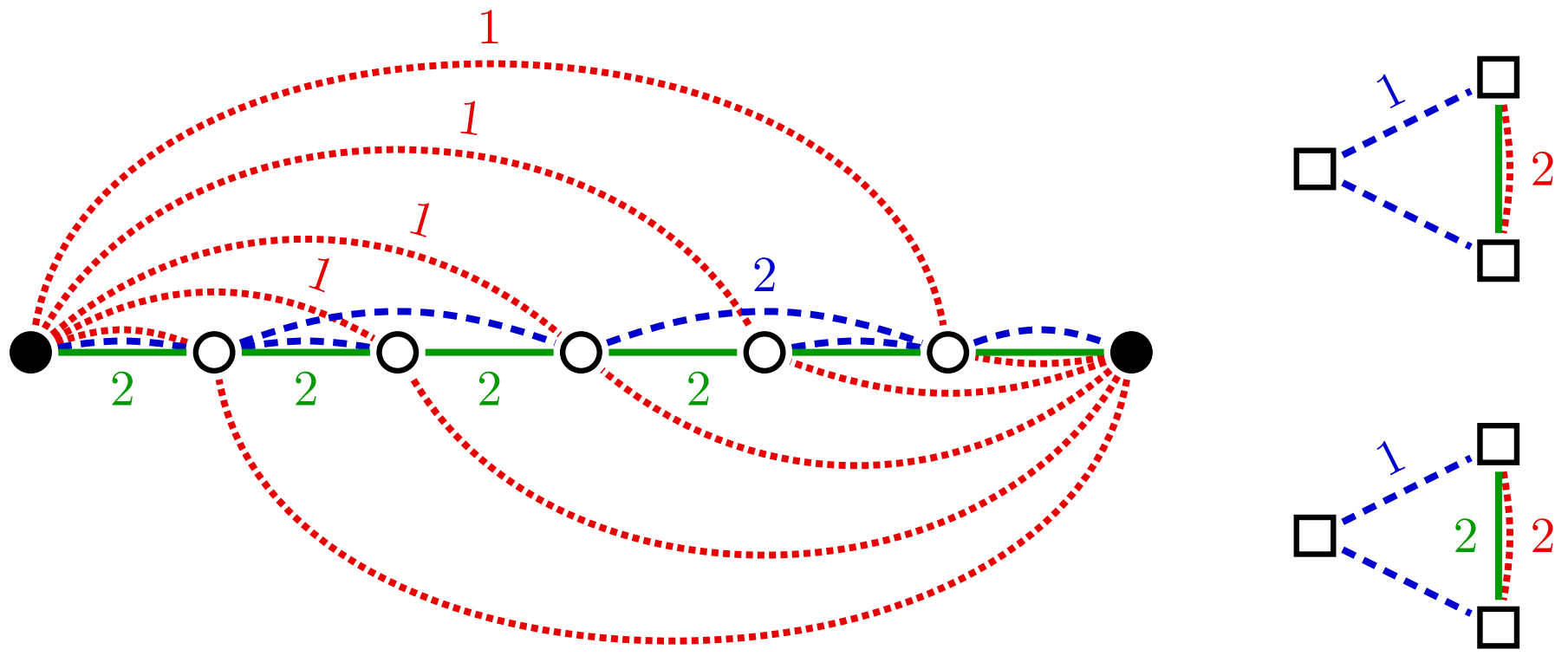

Our analysis of the algorithm is tight, as shown by the example in \Cref{fig:lp-tight}. 
(Here we assume that we use the Lau-Singh algorithm as a black-box.)

In the uniform special case, we obtain a better approximation factor.

\begin{theorem}
    \label{thm:unif-2-3}
    Let $\Delta \in \mathbb{N}$. If $b_v = \Delta$ for all $v\in V$, then \Cref{alg:nonuniformb} yields a 
    $\min\{ \frac{\Delta +1}{2}, \frac{\Delta + 1}{\Delta}\}$-approximation.
    In particular, the approximation factor is at most $\frac{3}{2}$.
\end{theorem}
\begin{proof}
    Recall that $F^\ast$ denotes a fixed optimal $\DBFs$ solution, and  $F', F''$ are the two solutions constructed by \Cref{alg:nonuniformb}.
    By \Cref{cor:w1-unif}, we have $w(F') \geq  \frac{\Delta}{\Delta +1 }\,w(F^\ast)$. 
    On the other hand, \Cref{lem:unif-w2} implies  $w(F'') \geq \frac{2}{\Delta +1}\, w(F^\ast)$. 
    Hence, returning a better solution among $F'$ and $F''$ yields an approximation factor of 
    \[
    \min\Big\{ \tfrac{\Delta +1}{2}, \tfrac{\Delta + 1}{\Delta}\Big\} \ \leq\ \tfrac{3}{2}. \qedhere
    \]
\end{proof}

\subsection{Integrality Gap Bound}\label{sec:integrality_gap}

    As a corollary of our algorithm, we show that the integrality gap of the natural LP relaxation with \emph{blossom constraints} is at most $\frac{8}{5}$. The formulation of the linear program is given below:
    \begin{equation}
        \tag{LP}
        \label{lp:lp}
        \begin{aligned}
        & \underset{x \in \mathbb{R}^{E}}{\text{maximize}}
        & \sum_{e\in E}w_e \,x_e \\
        & \text{s.t.} 
          &  x(E(S))            & \leq |S|-1 & \forall S\subseteq V,S \neq  \emptyset\\
        & & x(\delta(v))        & \leq b_v & \forall v\in V \\
        & & x(E(W)) + x(F)      & \leq  \left\lfloor \frac{b(W) + |F|}{2}\right\rfloor &  \substack{\forall W \subseteq V, F \in \delta(W) \\\mbox{ with } b(W) + |F| \mbox{ odd }} \\
        & & x_e                 &\in [0, 1] &\forall e\in E 
        \end{aligned}
    \end{equation}
    We define by $b(W) \coloneqq \sum_{v \in W} b_v$.
    The third type of constraints are blossom constraints. They were introduced by Edmonds \cite{Edmonds:1965:Maximum} and Pulleyblank \cite{Pulleyblank:2006:Facets} to describe the convex hull of integer solutions of the $b$-matching problem in (non-bipartite) graphs. 
    Blossom constraints are not necessary if the input graph is bipartite, in which case the \ref{lp:lp} is identical to the one used by Lau and Singh \cite{Singh-Lau:2015:Approximating}. 
    We refer to \ref{lp:lp}' as \ref{lp:lp} where blossom constraints are omitted. 
    We refer to \ref{lp:lp}'' as \ref{lp:lp} where the first constraints (bounding $x(E(S))$) are omitted. 
    In particular, \ref{lp:lp}'' is the standard $b$-matching linear program and is integral.
    
    Lau and Singh's iterative method produces a forest that might violate the degree bounds and whose cost is comparable to the cost of \ref{lp:lp}'. 
    \begin{theorem}[\cite{Singh-Lau:2015:Approximating}]
        \label{thm:LPapprox2}
        Given a graph $G=(V,E)$ with edge weights $w\colon E \to \mathbb{R}_{\geq 0}$ and degree bounds $b \colon V \to \mathbb{N}$, we can in polynomial time compute a forest $F$ such that 
        \begin{itemize}
            \item $w(F) \geq w(x^\ast)$, where $x^\ast$ is an optimal solution of \ref{lp:lp}' and
            \item  for each vertex $v \in V$, we have $\deg_{F}(v) \leq b_v + 1$.
        \end{itemize}
    \end{theorem}   
    As a corollary, we get the following. 
    \begin{corollary}
        \label{cor:integrality-gap}
        The integrality gap of \eqref{lp:lp} is at most $\frac{8}{5}$.
    \end{corollary}
    \begin{proof}
        Let $x^\ast$ be an optimal solution of \ref{lp:lp}. 
        Let $T',T''$ be the output of \textsc{Lau-Singh}$(G, w')$ and, \textsc{Max-Weight} $b$-\textsc{Matching}$(G, w'')$ and let $F', F''$ be the feasible degree-bounded forests computed by our algorithm (\Cref{alg:nonuniformb}).
        Observing that $x^\ast$ is a solution of \ref{lp:lp}', and applying \Cref{thm:LPapprox2}, the proof of \Cref{lem:w1} implies that
        \begin{align*}
            \sum_{e \in E} \frac{\bmin(e)}{\bmin(e) + 1}\, w_e\cdot x^\ast_e \ =\ w'(x^\ast) \ \leq\ w'(T') \ \leq\ w(F').
        \end{align*}
        Additionally, using that the $b$-matching polytope with blossom constraints is integral and that $x^\ast$ is a solution of \ref{lp:lp}'', the proof of \Cref{lem:w2} implies
        \begin{align*}
            \sum_{e \in E} \frac{1}{\bmin(e)}\, w_e \cdot x^\ast_e \ =\ w''(x^\ast) \ \leq \ w''(T'') \ \leq\ w(F'').
        \end{align*}
        Combining both inequalities as in the proof of \Cref{thm:5-8} yields $\max\{w(F'), w(F'')\} \geq \frac{5}{8} w(x^\ast)$.
    \end{proof}

\printbibliography

\end{document}